\documentstyle[12pt,a4]{article}
\begin{document}
\newcommand{\wst}{~{}^{{}^{*}}\llap{$\it w$}}
\newcommand{\wdst}{~~~{}^{{}^{**}}\llap{$\it w$}}
\newcommand{\omegast}{~{}^{{}^{*}}\llap{$\omega$}}
\newcommand{\omegadst}{~~~{}^{{}^{**}}\llap{$\omega$}}
\newcommand{\must}{~{}^{{}^{*}}\llap{$\mu$}}
\newcommand{\mudst}{~{}^{{}^{**}}\llap{$\mu$~}}
\newcommand{\nust}{~{}^{{}^{*}}\llap{$\nu$}}
\newcommand{\nudst}{~{}^{{}^{**}}\llap{$\nu$~}}
\newcommand{\beq}{\begin{equation}}
\newcommand{\eeq}{\end{equation}}
\newcommand{\gfrc}[2]{\mbox{$ {\textstyle{\frac{#1}{#2} }\displaystyle}$}} 
\title {\Large \bf   Bimetric Gravitation and Cosmology in Five Dimension  }
\author{{\normalsize \bf G S Khadekar} \thanks{Tel.91-0712-23946,
email:gkhadekar@yahoo.com} and {\normalsize \bf Bhavana Butey \thanks{ Department of Physics, G H Raisoni College of Engineering, Nagpur}} \\
{\small Department of Mathematics, Nagpur University} \\
{\small Mahatma Jyotiba Phule Educational Campus, Amravati Road}  \\
{\small Nagpur-440010 (INDIA) }}
\maketitle
\begin{abstract}
 Lee. et.al. (1976) analysed the bimetric theory with the help of parameterized post Newtonian (PPN) formalism. They found that the post Newtonian limit of the theory is identical with that of general theory of relativity except for their PPN parameter $(\alpha_{2})$ on the basis of cosmological considerations. In the present paper it is pointed out that feasibility of such considerations are doubtful in five dimensional bimetric theory of relativity.  As the universe is unique and is governed by physical laws, many different cosmologies are possible. Examples are given for some possible cosmological models, which are different, that those of Lee. et.al. This work is an extension in five dimension of a similar one obtained earlier by Rosen (1977) for  four-dimensional space-time.\\
\end{abstract}
\section { Introduction} 
Most recent efforts have been diverted at studying theories in which the dimensions of space-time are greater than (3+1) of the order which we observe. This idea is particularly important in the field of cosmology since one knows that our universe was much smaller in its early stage than it is today. Chados and Detweiler (1980) proposed the cosmological dimensional reduction process. They pointed out that the poor description of the universe is  obtained from Kasner's four-dimensional vacuum solutions since at least one-dimension contracts, whereas the other expands. This difficulty can be resolved with the introduction of the fifth dimension because the choice of the solution in this case is such that the five dimensional universes naturally evolves into an effect four dimensional one as a consequence of dimensional reduction. Higher dimensional space- time is particularly important in the domain of cosmology, in view of the underlying idea, that at its early stage of evolution our cosmos might have had a higher dimensional era such that with the passage of time the extra space reduced to volume beyond the ability of our experimental detection at the moment.
\par The bimetric general theory of relativity is a modification of Einstein's general relativity theory involving a background metric in addition to the usual physical metric. This theory is based on the assumption that at each point of the space-time there are a Riemannian metric tensor and a flat space metric tensor, satisfies the covariance and equivalence principles, as does general theory of relativity. The theory differs from general theory of relativity in that it does not appear to predict the existence of black holes and it gives a larger limit for the mass of a neutron star (Rosen \& Rosen 1975). PPN parameter of Rosen theory of gravitation is evaluated by Lee et. al (1976) and they showed that the post Newtonian limit of the theory is identical to that of general theory of relativity, except for the PPN parameter  $ \alpha_{2}$ . It follows that, in general,  $\alpha_{2} $ is different from zero. However,  according to  them, for a particular choice of cosmological boundary values, Newtonian parameter assumes its current value  $\alpha_{2} $ = 0, as in the case of general theory of relativity, and then the two theories are in complete agreement in the post Newtonian limit.
\par The present work is concerned with the cosmological considerations, which are doubtful in five dimensional (5D) Rosen theory of gravitation. Some examples are given for 5D cosmological model, which leads to conclusions that are different from those of Lee et.al (1976).
\subsection{An extreme example}
In bimetric theory there exit two metric tensors a Reimannian tensor $ g_{\mu\nu} $ describing the gravitational field (together with inertial forces) and a flat-space tensor $ \gamma_{\mu \nu}$ describing the inertial field, the two tensors being equal in the absence of gravitation. Accordingly, two kinds of covariant derivatives are defined : that involving    $ g_{\mu\nu} $, denoted by semicolon (;) , and that involving  $ \gamma_{\mu \nu}$ , denoted by $(\mid) $. The tensor $\gamma_{ \mu \nu}$ is denoted by the choice of the co-ordinate system ; the tensor  $ g_{\mu\nu} $ , by the field equations  are derived from the variational principle (Rosen 1973) and have the form (in the units of general relativity)
\beq
K _{\mu \nu} = -8 \pi T_{\mu \nu}.
\eeq
Here $ T_{\mu \nu} $ is the energy momentum density of matter or other non-gravitational  fields, and $ K_{\mu \nu} $
is given by 
\beq
\kappa K_{\mu \nu} = N_{\mu \nu} - \frac{1}{2} g_{\mu \nu} N,
\eeq
where 
\beq
\kappa = (\frac{g}{\gamma})^{\frac{1}{2}}, 
\eeq
\beq
N_{\mu \nu} =\frac{1}{2}\;\gamma^{\alpha \beta} g_{\mu  \nu {\mid _{\alpha \beta}}}  -  \frac{1}{2}\; \gamma^{\alpha \beta} g^{\lambda \sigma} g_{\lambda \mu {\mid_{\alpha }}}\; g_{\sigma \nu {\mid_{\beta}}},
\eeq
and
\beq
N = g^{\lambda \sigma} N_{\lambda \sigma}.
\eeq
From variational principal one obtains relations 
\beq
K^{\mu\nu}_{;\nu} = 0,
\eeq
\beq
T^{\mu \nu}_{;\nu} = 0.
\eeq
Unless otherwise indicated, indices are raised and lowered with $ g_{\mu \nu} $.
\par For describing homogeneous isotropic cosmological model let us take, in the rest frame of the universe, $ \gamma_{\mu\nu} = \eta_{\mu\nu} , $ the metric tensor of special relativity,  $ (diag. +1, -1, -1, -1, -1) ,$ and let us write the five dimensional line  element associated with $ g_{\mu \nu}, $
\beq
ds^2 = e^{2\phi} dt^2 - e^{2\psi} ( dx_{1}^2 + dx_{2}^2 + dx_{3}^2  + dx_{4}^2 ),
\eeq
where $ (t, x_{1},x_{2},x_{3},x_{4}) = ( x^0 ,  x^1, x^2, x^3, x^4), $ and $\phi = \phi(t), \psi = \psi(t).$ 
We are  dealing with model of zero spatial curvature $ (k = 0 ) $. If we take $ T^{\nu}_{\mu} $ as having non- vanishing components
\beq
T^{0}_{0} = \rho, \; \; \; T^{1}_{1}=T^{2}_{2}= T^{3}_{3} = T^{4}_{4} = - P,
\eeq
with  $\rho =\rho(t) $ and  $P = P(t) $, the density and the pressure, the field equations can be written as 
\beq
\ddot{\phi } = (\frac{-16 \pi}{3}) e^{(4\psi + \phi)} ( \rho +2P ),
\eeq
\beq
\ddot{\psi} =  (\frac{8 \pi}{3}) e^{(4\psi + \phi)} ( \rho - P ),
\eeq
where dot denotes a time derivative. In additions to these equations, from equation (7) are obtains a relation
\beq
\dot{\rho} + 4 ( \rho + P ) \dot{\psi } = 0.
\eeq
Consider the equation of state as
\beq
P = ( \gamma- 1) \rho \; \; \; \;\; (\gamma = const.).
\eeq
Let us take a case of pressureless dust $ ( \gamma  = 1) $. Then equation (12) can be written as 
\beq
\rho = \rho_{0} e^{-4\psi} \; \; \; \; \;  ( \rho_{0} = const.),
\eeq
and the field equation become
\beq
\ddot{\phi } = - (\frac{16}{3}) \pi {\rho_{0}} e^{\phi},
\eeq
\beq
\ddot{\psi} =  (\frac{8}{3})\pi  {\rho_{0}} e^{\phi}.
\eeq
The solution is given by 
\beq
e^{\phi} = (\frac{3{\delta}^{2}}{8{\pi} {\rho_{0}}})cosh^{-2}\delta(t-t_{0}),
\eeq
\beq
e^{\psi} = (\frac{8{\pi} {\rho_{0}}}{3{\delta}^{2}})^\frac{1}{2} e^\frac{At+B}{2} cosh\delta(t-t_{0}),
\eeq
where $ A, B, \rho_{0},\;\delta $ and $t_{0}$ are arbitrary constants.
For simplicity let us take $A= B = 0$ and let us introduce the cosmic time $\tau$ (Babala 1975),
\beq
\tau = \int\limits_{t_{0}}^{t}e^\phi dt,
\eeq
so that,
\beq
ds^2 = d\tau^2 - e^{2\psi} ( dx_{1}^2 + dx_{2}^2 + dx_{3}^2  + dx_{4}^2 ),
\eeq
we finds that,
\beq
\tau = (\frac{3{\delta}}{8{\pi} {\rho_{0}}}) tanh\delta(t-t_{0}),
\eeq
\beq
e^\psi =\left [{ \frac{3}{8{\pi}{\rho_{0}}({\tau_{0}}^2-{\tau^2})}}\right ]^{\frac{1}{2}} \;\;\;\;\;\; ( -{\tau_{0}} \leq {\tau} \leq {\tau_{0}}),
\eeq
with $\tau_{0} = (\frac{3{\delta}}{8{\pi} {\rho_{0}}})$. The density is given by 
\beq
\rho = (\frac{3\delta^2}{8\pi})^{2} \frac{1}{\rho_{0}\;cosh^4\delta(t-t_{0})}   =(\frac{8\pi}{3})^2  \rho_{0}^3\;(t_{0}^2 - t^2)^2.
\eeq
We see that this model describes a universe which contracts from a state with $ \rho = 0$ to one of maximum density $(\frac{3\delta^2}{8\pi})^{2} \frac{1}{\rho_{0}} $ and then expands again to the state with $ \rho = 0$. It should be remarked that the transformation from $ t $ to $ t_{0}$ which gives the line element (20) also changes the form of $\gamma_{\mu\nu}$. The above represents an example of the kind of cosmological model. One can take the tensor $g_{\mu\nu}$ for an isolated physical system as going over at infinity to the form corresponding to equation (8)
\beq
g_{00} =e^{2\phi} = c_0,\; \; \;  g_{jk} = - e^{2\psi} \delta_{jk} = - c_1\delta_{jk}.
\eeq
We assume that one of the no-zero PPN parameter in five dimensional bimetric theory of relativity in given by
\beq
\alpha_2 = (\frac{c_{0}}{c_{1}})-1.
\eeq
Lee.et.al. pointed out that, for a four dimensional cosmological model one can choose the boundary or (initial) conditions, so that $\alpha_2= 0$ (This can been seen in the above example, where one can make the expression (17) and (18) equal for an arbitrary valve of $t$ by a suitable choice of integration constant). Hence their consideration do not rule out the bimetric theory.
\par For example, let us suppose that in the homogenous isotropic model we have a unit vector $ S_\mu $ that points in the direction of flow of the cosmic time,
\beq
S_\mu =\frac{\partial\tau}{dx_\mu} \; ,\; S_\mu S^\mu =1,
\eeq
so that, in the co-ordinate system associated with equation (8)
\beq
S_0 = e^\phi ,S^0 = e^{-\phi} , S_k =S^k = 0 \;(k=1,2,3,4).
\eeq
Let us now assume that, in addition $T_{\mu\nu} $ which characterizes the matter, there also exist a ``cosmological" tensor $\Pi_{\mu\nu} $ which enters into the field equations, so that equation (1) is to be replaced by
\beq
K_{\mu\nu} = -8\pi \tau_{\mu\nu} +\Pi_{\mu\nu}.
\eeq
The tensor $\Pi_{\mu\nu} $ is to be regarded as given a prior. Let us assume that it has the form
\beq
\Pi_{\mu\nu} = 8\pi (\rho+P) (S_\mu S_\nu  -  \frac{1}{2}g_{\mu\nu}) - \Lambda g_{\mu\nu},
\eeq
where $\Lambda$ is constant. One of then finds that the equations (10) and (11) takes the form 
\beq
\ddot\psi=\ddot\phi =\frac{8\pi \kappa}{3}(\rho - P) + \frac{2}{3} \Lambda \kappa.
\eeq
With suitable initial conditions one has $\phi=\psi$. In this case equation (8) describes a conformly flat space-time and from (25) we see that $\alpha_2= 0$ for all time, as in general relativity.
Let us again consider a case of dust $ (P =0) $, so that equation (14) holds. With $\phi=\psi$ equation (30) gives
\beq
\ddot\psi =\frac{8\pi}{3}\rho_0 e^\psi +\frac{2}{3}\Lambda e^{5\psi}.
\eeq
Multiplying by $ \dot\psi$ and integrating , we gets
\beq
\frac{1}{2} {\dot\psi}^2 =\frac{8\pi}{3}\rho_0 e^\psi +\frac{2}{15} \Lambda e^{5\psi} +const.
\eeq
We see that, for $ \Lambda < 0$ and a suitable value  of the integration constant $(<0)$ , there will be two values of $\psi$ for which $\dot\psi =0$. Hence there will be solutions of equation (32) for which $\psi$ will oscillate with time between these values. We thus obtain an oscillating model of the universe.
\par However, it is possible to introduce other cosmic fields which will not violate the basic assumption of the theory and which can lead to models that are conformally flat over long periods of time. These will be considered in the next section. 
\subsection{ Cosmic Fields}
Rosen (1969) pointed out that, within the framework of general relativity theory, there may exist cosmic fields, which can influence the behaviour of the universe. These can be thought of as stress fields characterizing space-time. A similar possibility exists in the framework of the  bimetric theory. With the help of $g_{\mu \nu} $ and the cosmic tensor vector $ S_\mu $ given by (26), let us form a new tensor $ \Pi _{\mu \nu} $   which is  to appear on the right of equation (28). Now however,  in view of equation (6) and (7), let us require that
\beq
\Pi ^{\mu \nu}_{;\nu} = 0.
\eeq
One can take
\beq
\Pi_{\mu\nu} = \sigma (S_\mu S_\nu - \alpha g_{\mu \nu}) - \Lambda g_{\mu\nu},
\eeq
where $\sigma $ is a function of coordinates and $\alpha $ a constant. From  (26) we get 
\beq
 S_{\mu;\nu}S^{\nu} =0,
\eeq
so that equation (33) gives
\beq
(\sigma S^{\nu})_{;\nu} S_{\mu} - \alpha \sigma_{,\mu} =0 ,
\eeq
where comma denotes an ordinary partial derivative. Multiplying by  $ S^{\mu} $ we gets
$$[\sigma^{(1-\alpha)} S^{\nu}]_{; \nu} = 0 $$,
or
\beq
[((-1)^4g)^{\frac{1}{2}} \sigma^{(1-\alpha)} S^{\nu}]_{, \nu} =0.
\eeq
From equation (36) we also see that, if $ \alpha \ne 0 $, 
\beq
\sigma = \sigma(\tau).
\eeq
We can assume this also for $ \alpha =0$. In our coordinate system equation (37), with the help of (8), gives 
\beq
\sigma = \sigma_{0} e^{\frac {-4\psi}{(1-\alpha)}} \;\;\;\;\;\;\;\;\;\;\;\; (\sigma _0 = const.).
\eeq
If, formally, we write
\beq
\Pi_{\mu\nu}+{\Lambda} g_{\mu\nu} = (\rho_c  + P_c) {S_\mu}{ S_\nu} - {P_c}{ g_{\mu\nu}},
\eeq
as if we were dealing with a material medium having a velocity $ S^\mu $ ( $ \rho_c$ and $ P_c $ might be thought of something like density and pressure associated with empty space), them comparison with (34) gives
\beq
P_c = (\frac{\alpha}{1-\alpha}) \rho_c.
\eeq
The most interesting cases are : (1) $\alpha =0 , P_c =0, \sigma=\sigma_0e^{-4\psi}$, as in case of dust; (2)  $\alpha =  \frac {1}{4} , P_c = \frac{1}{3} \rho_c, \sigma =\sigma_0 e^\frac{-16}{3\psi}$, as for isotropic radiation, and (3) $\alpha =  \frac {1}{2} , P_c = \rho_c, \sigma =\sigma_0 e^{-8 \psi}$ , the limiting case, corresponding to a medium with acoustic velocity equal to light velocity.
\\
One can assume that any of the above fields (or those with other values of $\alpha$ )are present, either singly or combination. As an example, let us take as case $\alpha=0$.Then equation (28) take the form
\beq
K_{\mu\nu} = -8 \pi T_{\mu\nu} + \sigma S_\mu S_\nu -\Lambda g_{\mu\nu},
\eeq
which gives
\beq
\ddot\phi = - \frac{16}{3} \pi e^{4\psi +\phi} (\rho+2P) + \Lambda e^{4\psi +\phi} +\frac{2}{3} \sigma_0 e^\phi,
\eeq
\beq
\ddot\psi =\frac{8 \pi}{3} e^{4\psi + \phi} (\rho-P) + \Lambda e^{4\psi + \phi} - \frac{\sigma_0}{3} e^\phi.
\eeq
Let us again consider a case of dust-filled universe $(P=0)$. Making use of equation (14), we now get
\beq
\ddot\phi = - (  \frac{2}{3}) (8 \pi { \rho_0} - \sigma_0 ) e^\phi +\Lambda e^{(4 \psi +\phi)},
\eeq
\beq
\ddot\psi =   ( \frac{1}{3}) (8 \pi { \rho_0} - \sigma_0 ) e^\phi +\Lambda e^{(4 \psi +\phi)}.
\eeq
Here $\rho_0 $ and $\sigma_0 $ are integration constants. With suitable boundary condition we can have
\begin{center}
$ \sigma_0 = 8\pi \rho_0 $.
\end{center}
This value is similar to the value obtained earlier by Rosen (1977) for four dimensional space-time. From equations (45) and (46) gives
\beq
\ddot\phi = \ddot\psi =\Lambda e^{\phi+4\psi}.
\eeq
Again, with suitable boundary conditions, we can have $ \phi = \psi $, so that space-time is conformally flat, and our field equations becomes
\beq
\ddot\psi =\Lambda e^{5\psi}.
\eeq
Taking $ \Lambda >0 $ and choosing the integration constants appropriately, we get a solution of the form
\beq
e^\phi =e^\psi =(\frac{2 {\delta}^2}{5\Lambda})^\frac{1}{5} cosech ^ \frac{2}{5} (\delta t).
\eeq
We see that in the above example we again have a cosmological model which leads to  $\alpha_2= 0$, as in general relativity. The assumption that $P=0$, made above, is ~valid for the present and the future state of the universe; so if $\alpha_2 = 0 $ , at the present time, this will continue ~to hold for all time to come.
\subsection{Conclusion}
\par If one assumes that the universe is unique and is governed by a special law one can arrive at many different cosmological models, some examples of which are discussed above.
\bibliographystyle{plain}

\end{document}